\documentclass[ prd,tightenlines,nofootinbib,superscriptaddress,twocolumn]{revtex4-1}

\usepackage{hyperref}
\usepackage{graphicx}
\usepackage[utf8]{inputenc}
\usepackage[T2A,T1]{fontenc}	
\usepackage{textcomp}
\usepackage{amsmath}
\usepackage{amssymb}
\usepackage{amsthm}
\usepackage{amsfonts}
\usepackage{mathrsfs}
\usepackage{bbm}
\usepackage{tabularx}
\usepackage{multirow}	
\usepackage{eurosym}
\usepackage{color}
\usepackage[dvipsnames]{xcolor}
\usepackage{fancyhdr}
\usepackage{fancybox}
\usepackage{stmaryrd} 
\usepackage{slashed}
\usepackage{dsfont}
\usepackage{lastpage}
\usepackage{refcount}
\usepackage{CJKutf8}

\usepackage{todonotes}

\usepackage{tikz}
\usepackage{circuitikz}
\usetikzlibrary{shapes.geometric,calc,intersections,
decorations.markings,decorations.text,decorations,patterns, arrows,
decorations.pathmorphing,fpu,math,babel}

\newcommand{\me}{\mathrm{e}}

\newcommand{\md}{\mathrm{d}}

\DeclareMathOperator{\trace}{tr}

\newcommand{\ket}[1]{{\left|#1\right\rangle}}
\newcommand{\bra}[1]{{\left\langle#1\right|}}

\renewcommand{\C}{{\mathbb C}}
\newcommand{\N}{{\mathbb N}}

\newcommand{\Z}{{\mathbb Z}}

\newcommand{\cE}{{\mathcal E}}

\newcommand{\cS}{{\mathcal S}}

\newcommand{\SU}{\mathrm{SU}}

\newcommand{\be}{\begin{equation}}
\newcommand{\ee}{\end{equation}}
\newcommand{\beq}{\begin{eqnarray}}
\newcommand{\eeq}{\end{eqnarray}}
\newcommand{\bes}{\begin{eqnarray}}
\newcommand{\ees}{\end{eqnarray}}

\begin{document}

\title{Entanglement entropy and correlations in loop quantum gravity}

\author{{\bf Alexandre Feller}}\email{alexandre.feller@ens-lyon.fr}
\affiliation{Univ Lyon, Ens de Lyon, Université Claude Bernard Lyon 1, CNRS, 
Laboratoire de Physique, F-69342 Lyon, France}

\author{{\bf Etera R. Livine}}\email{etera.livine@ens-lyon.fr}
\affiliation{Univ Lyon, Ens de Lyon, Université Claude Bernard Lyon 1, CNRS, 
Laboratoire de Physique, F-69342 Lyon, France}

\begin{abstract}
Black hole entropy is one of the few windows toward the quantum aspects of gravitation
and its study over the years have highlighted the holographic nature of gravity. 
At the non-perturbative level in quantum gravity, promising explanations are being explored in terms of the entanglement entropy between regions of space. In the context of loop quantum gravity, this translates into the analysis of the correlations between regions of the spin network states defining the quantum state of geometry of space.
In this paper, we explore a class of states, motivated by results in condensed matter physics, satisfying an area law 
for entanglement entropy and having non-trivial correlations. We highlight that entanglement comes from holonomy operators acting on loops crossing the boundary of the region.
\end{abstract}

\maketitle

\section{Introduction} 

One of the most fascinating prediction of General Relativity, Einstein's classical theory of 
gravitation, is the existence of black holes, a region of space-time where nothing not even 
light can escape. Seen originally as a simple non-physical curiosity, the mathematical theory
of black holes is now a fully grown subject. A stationary black hole appears in fact to 
be quite a simple object defined entirely by its mass, angular momentum and charge (although this must be put in contrast with the recent work underlining the existence of classical hairs for black holes in relation to gravity's soft modes \cite{Strominger:2017aeh,Averin:2016hhm,Freidel_2016,Freidel_2016_2}). Moreover,
a straight analogy can be drawn between black holes dynamics predicted by Einstein's equation and 
thermodynamics. In particular, a notion of entropy is associated to a black hole \cite{Bekenstein_1973} which, in presence of quantum fields, is related to the area of the event horizon by the  Hawking formula \cite{Hawking_1975} and  has lead to the holographic principle relating geometric quantities and entropy in quantum gravity \cite{Bousso_2002}.

In the context of loop quantum gravity (see \cite{Rovelli_book,Rovelli_Vidotto_book,Thiemann_book}),
black hole entropy was mainly studied from the 
isolated horizon concept which assumes boundary conditions at the classical level 
\cite{Ashtekar_1998}. From a purely quantum perspective, this classical input 
should be removed. Instead, the quantum route is to compute an entanglement
entropy between a bipartite partition of the spin network state into an inside/outside
regions. For instance, in the 3d Riemannian BF formulation of gravity, such an 
entanglement (after a suitable regularization) satisfies an holographic behavior for 
the flat state \cite{Livine_Terno_2008} (see also \cite{Donnelly:2008vx,Donnelly:2011hn} for a more general treatment in loop gravity and lattice gauge theory beyond the flat state). Those calculations have strong similarities 
with those in some spin models like the toric code model useful for quantum computation
purposes \cite{Kitaev_2003,Zanardi_2005,Preskill_Kitaev_2006}. Here we would like to push further 
this similarities and propose to study a class of test states having holographic properties
and non trivial correlations. 

A very active line of research in this direction is tensor network renormalisation techniques applied to the context of the AdS/CFT correspondence for quantum gravity.  Indeed tensor network states turned out to be a tremendously efficient ansatz to study holography in quantum gravity \cite{Vidal_2015}, especially when looking at the holographic entanglement entropy \cite{Takayanagi_2006,Takayanagi_2009}. In particular, multi-scale entanglement renormalization ansatz (MERA) \cite{Vidal_2007} are especially promising as they appear to be variational ansatz of conformal field theory ground states and allow for a lattice realization of the AdS/CFT correspondence  \cite{Pastawski:2015qua,Hayden:2016bbf,Caputa:2017yrh}. Moreover, they have opened a fecund interaction between quantum gravity, quantum information and quantum computing.

Here, going in a similar direction although without using the MERA tools, our goal is to better understand the structure of correlations in loop quantum gravity (the interested reader can nevertheless find in \cite{Chirco:2017vhs} a first application of tensor network techniques to loop quantum gravity and spin network states). 
The motivation behind our study is twofold. The first one is to have a clearer understanding
of the physical states solving all the constraints of canonical quantum general relativity. It is expected
they should have non trivial correlations mapping to the two points correlation
functions of gravitons at the classical limit and that they should satisfy an area law.
We will introduce an ansatz for quantum states as superpositions of loop states with loops 
of arbitrary sizes with weights scaling for instance with the loop area, their perimeter
and their number. The second motivation is related to the definition and action of the Hamiltonian 
constraint of the theory which is still under active research. Identifying quantum states, with well-behaved correlations (both holographic and admitting nice 2-point correlations)
would give great insights toward the proper form and action of the quantum dynamic
implementing Einstein equation.
Our work can be seen as complementary to the study of entanglement on spin network states built from local Hamiltonian as in condensed matter models developed in \cite{Bianchi:2016hmk,Bianchi:2016tmw,Vidmar:2017uux}.

The present paper is structured as follows. Section \ref{toriccode_def} 
reviews the basic features of the toric code model that is then used to define 
the proper class of spin network states. The entanglement entropy between a partition 
(the system is a closed region) of the spin network is evaluated and we show that it scales as 
the number of degrees of freedom of the boundary in Section \ref{entanglement}. 
Loops crossing the boundary are seen 
to be at the origin of this entanglement. Section \ref{correlations} discusses a 
necessary generalization for the correlations to be non trivial and for the entanglement 
entropy to scale as the area. 

\section{The toric code model}
\label{toriccode_def}

The toric code model is a topological model of spin $1/2$ living on the links
of a general 2D lattice. The anyonic structure of the excitations makes it useful for 
fault-tolerant quantum computation \cite{Kitaev_2003,Zanardi_2005}. 
This model can be shown to be equivalent to a $BF$ theory on the discrete 
group $\Z_2$, a highly interesting fact since gravity can be formulated as a 
(constrained) $BF$ theory.

Considering a  bipartite partition of the lattice, it was found that the entanglement 
entropy between those regions were proportional to the boundary area 
(plus a topological term) which is reminiscent of the holographic principle. 
We will review here the basic results useful for our following discussion 
on spin network states.

Let's define $n_s$ and $n_p$ the number of vertex and plaquettes respectively.
The dynamic of the model is constructed with the vertex operators 
$A_s = \bigotimes_{j\in s} \sigma_j^x$, tensor product of Pauli matrices with the 
vertex $s$ as a source, and plaquette operators 
$B_p = \bigotimes_{j\in\partial p} \sigma_j^z$, tensor product of Pauli matrices 
around the plaquette $p$. The Hamiltonian is then 
\begin{align}
H = - \sum_{s=1}^{n_s} A_s - \sum_{p=1}^{n_p} B_p
\end{align}
We stress here the fact that the plaquette and vertex operators are subject
to a particular constraint $\prod_p B_p = \prod_s A_s = \mathbbm{1}$. 
Every term commute with all the others which makes it easier to find the ground
state(s) $\ket{\psi_0}$ of the system by looking at the fundamental of each operators (states 
diagonalizing all the operators with highest eigenvalues)
\begin{align}
A_s\ket{\psi_0} = B_p\ket{\psi_0} = \ket{\psi_0}
\end{align}
The lowest energy state of the vertex operators can be seen as gas of loops.
It  implements the Gauss law enforcing gauge invariance at each vertex. 
Taking into account the plaquette operators, which deform  smoothly a loop into another, 
the ground state $\ket{\psi_0}$ of the system is a superposition of all
those loops with equal weights and imposes the flatness of the $\Z_2$ holonomy.
Denoting by $\mathcal{A}$ the group generated by the vertex operators, we have
\begin{align}
\ket{\psi_0} = \frac{1}{\sqrt{2^{n_s -1}}} \sum_{g\in\mathcal{A}}g\ket{0}
\end{align}
In fact, because a plaquette operator only smoothly deform a loop, the fundamental
subspace is degenerate, with dimension $2^{2g}$ for an orientable genus $g$ 
surface. This degeneracy is at the heart of the topological character of this model 
which cannot be lifted by local perturbations.

To cast thing in the proper form useful for generalizing to gravity (for the 
$\SU(2)$ group), we write explicitly the expanded form of the ground states
in terms of loops. Written in terms of projector $\left( \mathbbm{1} + B_p \right) / \sqrt{2}$, 
$
\ket{\psi_0} = \frac{1}{\sqrt{2^{n_p +1}}} \prod_p \left( \mathbbm{1} + B_p \right) \ket{0}
$,
the loop expansion is straightforward. It simply suffices to 
expand the product of operators. Thanks to the fact that $\sigma_i^2 = \mathbbm{1}$,
the product of two plaquette operators $B_{p_1}$ and $B_{p_2}$ sharing one
link is equivalent to an operator on the disjoint union of the plaquettes 
$B_{\partial (p_1 \cup p_2)}$. We have finally the loop superposition form
\begin{align}
\label{toriccode_loopstate}
\ket{\psi_0} =\frac{1}{\sqrt{2^{n_p -1}}} \sum_{\mathcal{C}} 
\bigotimes_{{\mathcal{L}\in\mathcal{C}}}\ket{1_{e\in\mathcal{L}}, 0_{e\not\in\mathcal{L}}}
\end{align}
Here the set $\mathcal{C}$ is the set of all configuration of non intersecting 
loops having no links in common.

\begin{figure}[h]
  \centering

\begin{tikzpicture}[domain=-2.2:2.2]

\tikzstyle{dot}=[draw,circle,minimum size=2pt,inner sep=0pt,outer sep=0pt,fill=black]

	\def\a{.6}
	\def\nbh{5}
	\def\nbw{6}
	\pgfmathsetmacro{\l}{0.5*\a}
	
	\draw[blue, dashed] (1.5*\a,1.5*\a) -- (4.5*\a,1.5*\a) -- (4.5*\a,3.5*\a) -- (1.5*\a,3.5*\a) -- cycle ;
	\node at (4.5*\a,0.5*\a) {\textcolor{blue}{$\mathcal{SE}$}};

	\draw (0,0) -- (\a,0) -- (\a,\a) -- (0,\a) -- cycle;
	\node at (0.5*\a,4.5*\a) {$\mathcal{E}$};
	
	\draw[red, thick] (\a,4*\a) -- (\a,\a) -- (2*\a,\a) -- (2*\a,0) -- (3*\a,0) -- (3*\a,2*\a) -- (2*\a,2*\a) -- (2*\a,4*\a) -- cycle ;
	\draw[red, thick] (3*\a,4*\a) -- (5*\a,4*\a) -- (5*\a,2*\a) -- (4*\a,2*\a) -- (4*\a,3*\a) -- (3*\a,3*\a) -- cycle ;
	\node at (3.5*\a,2.5*\a) {$\mathcal{S}$};

	\foreach \i in {0,...,\nbw} {
		\foreach \j in {0,...,\nbh} {
					\coordinate[dot, black] () at (\i*\a,\j*\a);
		}
	
}

\end{tikzpicture}

  \caption{Illustration of one possible loop structure appearing in the superposition 
  		defining the Kitaev state motivated by the ground state structure of the 
  		toric code model. Three kind of loops are distinguished when a subsystem 
  		is chosen and only loops crossing the boundary give a non zero entanglement.}
  \label{loops}
\end{figure}
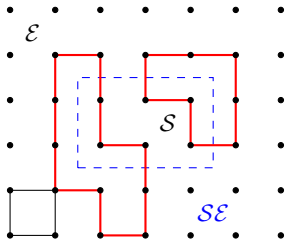

As mentioned above, one of the interesting result of the toric code model 
is the area scaling law of the entanglement entropy for the ground state
between a bipartite partition of the lattice \cite{Zanardi_2005}. 
Consider a region $S$ and its exterior $E$, the global system being in a 
ground state, see Fig.\ref{loops}. 
The reduced density matrix of the region $S$ is needed to obtain the entropy. 
The loop structure coming from the state \eqref{toriccode_loopstate} is composed 
of three kinds of loops, those contained completely in $S$ or $E$ and those
belonging to both. We then have the following 
result (with a detailled proof in annex \ref{proofs}):
the entanglement entropy associated to a 
given region $S$ whose (contractible) frontier possesses $n_{SE}$ degrees
of freedom in the ground state is 
\begin{align}
S = n_{SE} - 1
\end{align}
This entropy is 
proportional to the number of degrees of freedom on  the boundary and 
scales as the area. The minus one is a topological contribution and is model
dependent in some sense \cite{Preskill_Kitaev_2006}.

What we intend to do now is the study in the context of loop
quantum gravity the class of wavefunction 
having the same loop structure as the one for the toric code
model and study the entanglement entropy and correlations they 
contain.

\section{Definition and properties}
\label{spinnetstate_def}

\subsection{The loop decomposition}

The holographic principle is one of the few accepted feature every
quantum theory of gravity should have. Simply stated, it says that
volume degrees of freedom of a region of spacetime are encoded
on some degrees of freedom on the boundary \cite{Bousso_2002}.
 We saw that the entanglement entropy of the toric 
code model has this same behavior. The purpose here is then 
quite simple: we want to adapt the ground state structure of this
model for the spin network state on a given  random 2d lattice 
with edge degrees of freedom fixed to the fundamental spin $1/2$
excitation (this condition can be relaxed by choosing any spin $j$).

The natural $\SU(2)$ gauge invariant object in loop quantum gravity 
is the holonomy, here 
$\chi_{1/2}	\left(\prod_{e\in \mathcal{L}} g_e \right) $
for a given loop $\mathcal{L}$ and group element $g_e \in \SU(2)$ 
for each edge. We thus define by analogy the state which has the same loop structure 
than \eqref{toriccode_loopstate}. In fact, canonical model of statistical physics
such as the Ising model or $O(N)$ models, hints toward adding new amplitude
contribution like a perimeter $P(\mathcal{C})$ contribution $\gamma^{P(\mathcal{C})}$ 
or/and a  number of loops $N(\mathcal{C})$ contribution  $ \beta^{N(\mathcal{C})}$.
So the natural general states we are interested in are given by the wave function
\begin{align}
\label{def_state}
\psi_{\alpha, \beta, \gamma}(g_e) = 
	\sum_{\mathcal{C}} \alpha^{A(\mathcal{C})} \beta^{N(\mathcal{C})} \gamma^{P(\mathcal{C})}
		\prod_{\mathcal{L}\in\mathcal{C}}
			\chi_{1/2}\left(\prod_{e\in \mathcal{L}}^{\rightarrow} g_e \right) 
\end{align}
A given configuration $\mathcal{C}$ is composed of non intersecting
loops $\mathcal{L}$ having no links in common while $A(\mathcal{C}) $,
$N(\mathcal{C})$ and $P(\mathcal{C})$ are respectively the total area, the 
number of loop and the perimeter of the configuration and $\alpha,\beta,
\gamma \in \C$  are complex amplitudes. 

Our goal is to study this class of states, the scaling law of the entanglement
entropy between a partition of the spin network and then the correlation
two point functions between spins of different edges. We start with the 
simple case $\beta=\gamma=1$ and see that the
entropy scales as expected with the number of degrees of freedom of the 
boundary. However, the correlations will appear to be topological,
 motivating the introduction of a more general class of states with 
amplitudes function of the perimeter of the loops or their number.  

In fact, we could have first thought of a simpler state constructed as a product
of all holonomies of each plaquette (with a potential contribution from a boundary
for a finite size graph) as 
\begin{align*}
\psi(g_e) = \prod_p \chi_{1/2}\left(\prod_{e\in p}^{\rightarrow} g_e \right) 
			\chi_{1/2}\left(\prod_{e\in \partial}^{\rightarrow} g_e \right)
\end{align*}
At first sight, it would appear that such a state would display some non trivial 
correlations. However this is not the case both for the holonomy and spin two
point functions in the infinite size limit. We won''t dwell on this state in the 
core of this paper, see annex \ref{naive} for more details.

%

\subsection{Behavior under coarse-graining}

The state $\psi_{\alpha, \beta, \gamma}$ have very nice properties under
some coarse-graining procedures due the very particular loop structure we 
chose. For a graph $\Gamma$, one procedure is to simply eliminate a link $e_0$
constructing the new graph $\Gamma\setminus e_0$ and another is to 
pinch the link to a node defining the pinched graph $\Gamma - e_0$.

From the wave function $\psi_{\alpha, \beta, \gamma}^\Gamma$, the pure 
elimination of a link is done by a simple average. Here, since the 
loops composing the state are always non overlapping, 
the integration over $e_0$ amounts to remove all loops containing it.
\begin{align}
\int_{\SU(2)}\psi_{\alpha, \beta, \gamma}^\Gamma \left(g_{e_0},g_e\right) \; \md g_{e_0}
=\psi_{\alpha, \beta, \gamma}^{\Gamma\setminus e_0}(g_e)
\end{align}
Thus the coarse-grained state corresponds exactly to the state on the coarse-grained 
graph $\Gamma\setminus e_0$. We have a stability under this coarse-graining procedure. 

\begin{figure}[h]
  \centering

\begin{tikzpicture}[domain=-2.2:2.2]

\tikzstyle{dot}=[draw,circle,minimum size=2pt,inner sep=0pt,outer sep=0pt,fill=black]

	\def\a{1}
	\def\nbh{1}
	\def\nbw{2}
	
		\matrix[] () at (0,0) {
	
	\draw (\a,0) -- (0,0) -- (0,\a) -- (\a,\a) ;
	\node at (-0.3*\a,0.5*\a) {\textcolor{blue}{$h_S$}};

	\draw (\a,0) -- (2*\a,0) -- (2*\a,\a) -- (\a,\a) ;
	\node at (2.3*\a,0.5*\a) {$h_E$};
	
	\draw[red, thick] (\a,0) -- (\a,\a) ;
	\node at (1.2*\a,0.5*\a) {$h_b$};

	\foreach \i in {0,...,\nbw} {
		\foreach \j in {0,...,\nbh} {
					\coordinate[dot, black] () at (\i*\a,\j*\a);}}

		&

		\draw [->] (-\a/3,\a/2) -- (\a/3,\a/2)  ;
		\node at (0,0.3*\a) {$h_b=\mathbbm{1}$};

		&
		
	\draw (\a,\a/2) -- (0,0) -- (0,\a) -- cycle ;
	\node at (-0.3*\a,0.5*\a) {\textcolor{blue}{$h_S$}};

	\draw (\a,\a/2) -- (2*\a,0) -- (2*\a,\a) -- cycle ;
	\node at (2.3*\a,0.5*\a) {$h_E$};
	
	\draw[dotted] (0,0) -- (2*\a,0);
	\draw[dotted] (0,\a) -- (2*\a,\a);
	\draw[dotted] (\a,\a) -- (\a,0);

	\coordinate[dot, black] () at (0,0);	
	\coordinate[dot, black] () at (0,\a);
	\coordinate[dot, black] () at (2*\a,0);
	\coordinate[dot, black] () at (2*\a,\a);
	\coordinate[dot, red] () at (\a,\a/2);  \\
	};

\end{tikzpicture}

  \caption{The pinch coarse-graining method is an invariant procedure only for 
  the case $\gamma = 1$, meaning the state doesn't contain perimeter information.}
  \label{example}
\end{figure}
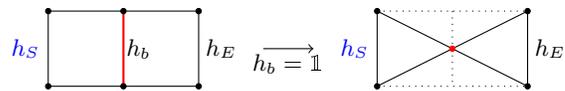

The second method is to pinch the link. This is done by imposing the holonomy 
on $e_0$ to be equal to the identity. The coarse-grained state is $\left.
\psi_{\alpha, \beta, \gamma}^\Gamma(g_e)\right|_{g_{e_0} = \mathbbm{1}}$.
For a given loop containing $e_0$, pinching the link doesn't change the area or
the number of loops, but only its perimeter. Separating configuration containing 
the link $e_0$ or not, forming respectively the sets $\mathcal{C}_0$ 
and $\mathcal{C}\setminus e_0$, we have
\begin{align}
\left.\psi_{\alpha, \beta, \gamma}^\Gamma(g_e)\right|_{g_{e_0} = \mathbbm{1}} &=
	\sum_{\mathcal{C}\setminus e_0}
		\alpha^{A(\mathcal{C})} \gamma^{P(\mathcal{C})}
		\prod_{\mathcal{L}\in\mathcal{C}}
			\beta\chi_{1/2}\left(\prod_{e\in \mathcal{L}}^{\rightarrow} g_e \right)  \nonumber \\
	&+\gamma \sum_{\mathcal{C}_0}
		\alpha^{A(\mathcal{C})} \gamma^{P(\mathcal{C})}
		\prod_{\mathcal{L}\in\mathcal{C}}
			\beta\chi_{1/2}\left(\prod_{e\in \mathcal{L}}^{\rightarrow} g_e \right) \nonumber
\end{align}
The invariance under coarse-graining is recovered at the condition that $\gamma = 1$, 
meaning that the perimeter of the loops doesn't matter: 
$\left.\psi_{\alpha, \beta, \gamma}^\Gamma(g_e)\right|_{g_{e_0} = \mathbbm{1}} =
\psi_{\alpha, \beta, \gamma}^{\Gamma-e_0} (g_e)$.

\section{Entanglement entropy}
\label{entanglement}

\subsection{Entanglement entropy}

The next step is to compute the entanglement (Von Neuman) entropy 
$S= \trace{(\rho_{\cS} \ln\rho_{\cS})}$ between a bipartite partition of the graph.
The system $\cS$ of interest will be a bounded connected region and 
the rest of the graph forms the environment $\cE$ whose degrees of 
freedom are traced out. The boundary group elements will be by convention 
incorporated into the system and won't be traced over.
For simplicity, we will restrict the evaluation of the entropy for $\beta= \gamma = 1$.


To compute the entropy of $\cS$, we need its reduced density matrix defined as 
 \begin{align}
\rho_{\cS}(\tilde{g}_e, g_e) = \int
	\overline{\psi}_{\alpha}(\tilde{g}_{e\in\cS}, h_{e\notin \cS}) 
	\psi_{\alpha}(h_{e\notin \cS}, g_{e\in\cS}) \; \md h_{e\notin \cS}
\end{align}
The method to evaluate the entropy $S= -\trace{(\rho_{\cS} \ln\rho_{\cS})}$
is based on the replica trick \cite{Wilczek_1994}. Computing the successive power of the reduced 
density matrix $\rho_{\cS}^n, \; n\in\N$, we then obtain the entropy 
by $S = - \left.\frac{\partial \trace{\rho_\cS^n}}{\partial n}\right|_{n=1}$.

The first step is to compute the reduced density matrix.
Denoting respectively $\mathcal{C}_S$, $\mathcal{C}_E$ and $\mathcal{C}_{SE}$ 
the loops belonging to $S$, $E$ or both, we have (see annex \ref{Entanglement Kitaev state})
\begin{widetext}
\begin{align}
\label{reduced_density}
\rho_{S}(g,g') = \frac{\mathcal{N}_E (\alpha)}{\mathcal{N}(\alpha)} 
		\sum_{\substack{
				\mathcal{C}_S \cup \mathcal{C}_{SE} \\
				\mathcal{C}'_S \cup \mathcal{C}_{SE}
					}
			}
			\overline{\alpha}^{A(\mathcal{C}'_S \cup \mathcal{C}_{SE})} 
			 \alpha^{A(\mathcal{C}_S \cup \mathcal{C}_{SE})} 
			 \times \prod_{\substack{
				\mathcal{L}_S\in\mathcal{C}_S \\
				\mathcal{L}'_S\in \mathcal{C}'_S
					}
			}
			\chi_{1/2}\left( \mathcal{L}_S (g)\right)
			\chi_{1/2}\left( \mathcal{L}'_S (g')\right)
		\prod_{\mathcal{L}_{SE}\in\mathcal{C}_{SE}}
		\left(
		\frac{1}{2} 
			\chi_{1/2}\left( \mathcal{L}_{SE} (g,g') \right)
		\right)
\end{align}
\end{widetext}
with $\mathcal{N}(\alpha)$ the norm and $\mathcal{N}_E (\alpha) = 
\sum_{\mathcal{C}_E} |\alpha|^{2A(\mathcal{C}_E)}$ the factor coming 
out of the partial trace on the environment. We see that two contributions 
appear, one with only loops in $S$ and another coming from loops crossing
the boundary. This last term is responsible for the entanglement between
$S$ and $E$. 

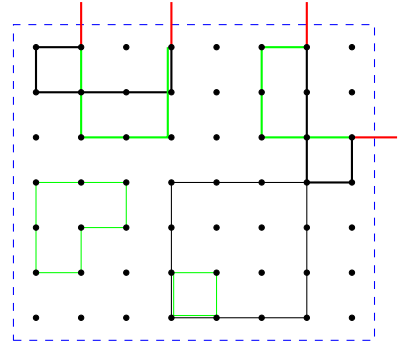
\begin{figure}[h]
  \centering

\begin{tikzpicture}[domain=-2.2:2.2, rotate=-90]

\tikzstyle{dot}=[draw,circle,minimum size=2pt,inner sep=0pt,outer sep=0pt,fill=black]

	\def\a{.6}
	\def\nbh{7}
	\def\nbw{6}
	\pgfmathsetmacro{\l}{0.5*\a}

	\draw[blue, dashed] (-0.5*\a, -0.5*\a) -- (-0.5*\a,7.5*\a) -- (6.5*\a,7.5*\a) -- (6.5*\a,-0.5*\a) -- cycle;
	
	\draw[red,thick] (-\a,\a) -- (0,\a) ;
	\draw[red,thick] (-\a,3*\a) -- (0,3*\a) ;
	\draw[red,thick] (-\a,6*\a) -- (0,6*\a) ;
	\draw[red,thick] (2*\a,8*\a) -- (2*\a,7*\a) ;
	
	\draw[black,thick] (0,\a) -- (0,0) -- (\a,0) -- (\a,3*\a) -- (0,3*\a) ;
	\draw[black,thick] (0,6*\a) -- (3*\a,6*\a) -- (3*\a,7*\a)-- (2*\a,7*\a) ;
	\draw[black] (3*\a,6*\a) -- (6*\a,6*\a) -- (6*\a, 3*\a) -- (3*\a,3*\a) -- cycle ;
	
	\draw[green,thick] (0,\a) -- (2*\a,\a) -- (2*\a,2.92*\a) -- (0,2.92*\a) ;
	\draw[green] (3*\a,2*\a) -- (3*\a,0) -- (5*\a,0) -- (5*\a,\a) -- (4*\a,\a) -- (4*\a,2*\a) -- cycle ;
	\draw[green] (5*\a,3.05*\a) -- (5.95*\a,3.05*\a) -- (5.95*\a,4*\a) --(5*\a,4*\a) -- cycle ;
	\draw[green,thick] (0,6*\a) -- (0,5*\a) -- (2*\a,5*\a) -- (2*\a,7*\a) ;

	\foreach \i in {0,...,\nbw} {
		\foreach \j in {0,...,\nbh} {
					\coordinate[dot, black] () at (\i*\a,\j*\a);
		}
	
}

\end{tikzpicture}

  \caption{Illustration of one possible loop structure after the partial trace over 
  the environment has been performed. The reduced density matrix is not factorized 
  anymore and a non trivial boundary contribution leads to entanglement.}
  \label{loops_reduced}
\end{figure}

Figure \ref{loops_reduced} shows an example of a configuration appearing 
in the reduced density matrix. To understand simply the form of $\rho_\cS$, let's imagine
we have only two loops configuration, one copy for the bra and ket of 
the density matrix. Each configuration is composed of non overlapping and 
non intersecting loops. Nonetheless, each copy can overlap since their are independent.
Now, tracing out the $E$ degrees of freedom imposes that the parts in $E$ from 
each copies to be exactly the same, otherwise the average
gives zero. Complications come from loop crossing the boundary.
The average of the $\cE$ part of crossing loops gives a contribution of the from
$\int_{\SU(2)} \chi_{1/2}(gh) \chi_{1/2}(g'h) \; \md h
= \frac{1}{2} \chi_{1/2}(gg'^{-1})$.
This is at the origin of the boundary holonomies in \eqref{reduced_density}.
Now considering again all the allowed configurations, we see that for a given bulk/boundary 
plaquette choice like in Fig.\ref{loops_reduced}, their is a huge redundancy
coming from the $E$ plaquettes. After the partial trace, this leads to the overall
$\mathcal{N}_E (\alpha)$ prefactor.

The next step is to compute the successive power and take
the trace. At the end a simple formula remains, 
\begin{align*}
\trace{\rho_{S}^n (g,g')}&=\left( \frac{\mathcal{N}_E (\alpha)}{\mathcal{N}(\alpha)} \right)^{n}
				\mathcal{N}^{n-1}_S (\alpha)
		\sum_{\mathcal{C}_S \cup \mathcal{C}_{SE}} 
			\frac{|\alpha^{A(\mathcal{C}_S \cup \mathcal{C}_{SE})}|^{2n}}
				{\left(4^{n-1}\right)^{\#\mathcal{C}_{SE}}}
\end{align*}
Appendix \ref{proofs} presents the detailed calculations. The entropy 
of the region $\cS$ is then directly obtained by differentiation and we have that 
the leading order term scales as the area of the boundary of the region. The
entanglement  entropy is given finally by
\begin{align}
\label{entropy}
S =	n_{SE} f(|\alpha|^2) +
	\frac{2\ln 2}{\left( 1 + |\alpha|^2 \right)^{n_{SE}}} 
		\sum_{\mathcal{C}_{SE}}\#\mathcal{L}_{SE} |\alpha|^{2A(\mathcal{L}_{SE})} 
\end{align}
with $n_{SE}$ the number of degrees of freedom at the boundary 
and $f(|\alpha|^2) = \ln\left( 1 + |\alpha|^2 \right) - 
\frac{|\alpha|^2 }{ 1+ |\alpha|^2 } \ln\left( |\alpha|^2 \right) $.
This formula is quite general and is valid for an arbitrary graph as long as the 
loop structure of the spin network state is the same.

\subsection{Boundary degrees of freedom - Purification}

We came to understand that the entanglement in the subsystem $\mathcal{S}$
prepared in the state \eqref{reduced_density} can be traced back to loops
crossing the boundary. In fact, we can understand the state \eqref{reduced_density} 
as resulting from tracing out additional boundary degrees of freedom. This idea
goes in the same spirit as recent studies on local subsystems in gauge theories and 
gravity \cite{Freidel_2016,Freidel_2016_2}.

To purify the state, consider at each puncture a new degree of freedom, for 
instance a new fictitious edge. We work in the extended Hilbert space 
$\mathcal{H}_{\mathcal{S}} \otimes \mathcal{H}_e^{\otimes N}$ with $\mathcal{H}_e$
the Hilbert space associated to the new edge, $N$ the total number 
of puncture and $\mathcal{H}_{\mathcal{S}}$ the Hilbert space of the system. 
We then construct a pure state as superposition of loops in the bulk and 
paths joining pairs of punctures, see for instance FIG.\ref{loops_reduced}.
For a path $\mathcal{P}$, we use the holonomy (properly oriented)
\begin{align}
\chi_{1/2}\left(\mathcal{P}\right) = \chi_{1/2}(h_{s_b}g_{\mathcal{S}}h_{t_b})
\end{align}
with $h_{s_b,t_b}$ associated to the pair of boundary degrees of freedom, the source
and target of the path respectively. The reduced density matrix \eqref{reduced_density}
can then be purified by considering the state $\ket{\psi_{\mathcal{S}B}}$ (B for boundary)
with wave-function
\begin{align}
\psi_{\mathcal{S}B}(g) = 
	\sum_{\mathcal{C}} \alpha^{A(\mathcal{C})} \beta^{N(\mathcal{C})} \gamma^{P(\mathcal{C})} 
		\prod_{\mathcal{L}\in\mathcal{C}}
			\chi_{1/2}\left(\mathcal{L} \right) 
		\prod_{\mathcal{P}\in\mathcal{C}}
			\chi_{1/2}\left(\mathcal{P} \right) 
\end{align}
Then $\rho_\mathcal{S} = \trace{\left( \ket{\psi_{\mathcal{S}B}}\bra{\psi_{\mathcal{S}B}} \right)}$.
We have purified the reduced density matrix of the system. One could argue that their are
many ways to purify a quantum state and could question its relevance here. After all the original
state \eqref{def_state} is a perfectly valid purification. What is really interesting here is the method. 
We can think of the local subsystem on its own by doubling the boundary degrees of freedom 
and construct pure state in an extended Hilbert space. The physical state is recovered by 
tracing out the additional boundary degrees of freedom. This match exactly the results of
\cite{Freidel_2016} by a direct analysis of the reduced density matrix of a sub-region 
of the spin network state. Naturally we here have no particular information on those
additional degrees of freedom and they should be determined by a proper analysis of
boundary terms in the classical and quantum theory.

This elementary discussion illustrates simply the fact that the extended Hilbert space
method can been seen from a quantum information perspective as a clever way 
to purify a state of a local region and consequently why it has something to say 
about entanglement, correlations and entropy.

\subsection{On correlations}

This holographic behavior is a good sign for this class of states to be good candidates 
for physical states solutions of the Hamiltonian constraint of loop quantum gravity. 
What's more, for physical solutions, we expect the correlations between geometrical
observables to be non trivial. This is where the limit $\beta= \gamma = 1$ fails.
Indeed, the spin (or holonomies) two points correlation functions are topological in the 
sense that they do not depend on the graph distance between the edges. 

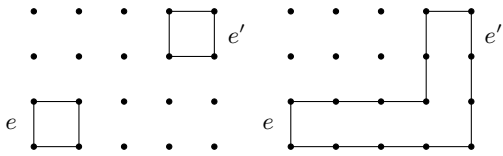
\begin{figure}[h]
  \centering

  \begin{tikzpicture}[domain=-2.2:2.2]

\tikzstyle{dot}=[draw,circle,minimum size=2pt,inner sep=0pt,outer sep=0pt,fill=black]

	\def\a{.6}
	\def\nbh{3}
	\def\nbw{4}
	\pgfmathsetmacro{\l}{0.5*\a}
	
	\matrix[] () at (0,0) {
	
	\draw (0,0) -- (\a,0) -- (\a,\a) -- (0,\a) -- cycle;
	\draw (3*\a,2*\a) -- (3*\a,3*\a) -- (4*\a,3*\a) -- (4*\a,2*\a) -- cycle;
	\node at (-0.5*\a,0.5*\a) {$e$};
	\node at (4.5*\a,2.5*\a) {$e'$};

	\foreach \i in {0,...,\nbw} {
		\foreach \j in {0,...,\nbh} {
					\coordinate[dot, black] () at (\i*\a,\j*\a);
		}
	
	}
	
		&
		\hspace{-0.5cm}
		&

	\draw (0,0) -- (4*\a,0) -- (4*\a,3*\a) -- (3*\a,3*\a) -- (3*\a,\a) -- (0,\a) -- cycle;
	\node at (-0.5*\a,0.5*\a) {$e$};
	\node at (4.5*\a,2.5*\a) {$e'$};

	\foreach \i in {0,...,\nbw} {
		\foreach \j in {0,...,\nbh} {
					\coordinate[dot, black] () at (\i*\a,\j*\a);
		}
	
	}
 \\
	};

\end{tikzpicture}

  \caption{Trivial correlations arise because their is no distinction between configurations
  presented is the figure.}
  \label{correlations_example}
\end{figure}

Let's look for instance at the spin two points functions 
$\langle \hat{j}_e \hat{j}_{e'} \rangle - \langle \hat{j}_e \rangle \langle \hat{j}_{e'} \rangle$. 
This spin operator is defined by its action on a spin network state with the help
of the Peter-Weyl theorem $(\hat{j}_e \psi)(g,g_e) = \sum_{j_e} (2 j_e +1) j_e
\int \chi_{j_e}(g_eh^{-1})\psi(g,h) \;\md h$. The method to evaluate the averages 
goes as follows. First we have only the spin $1/2$ component of the average that 
gives a non zero contribution, so that we have
\begin{align*}
\langle \hat{j}_e \rangle  
&= \frac{	\int_{SU(2)}\! 
		\chi_{1/2}(g_eh_e^{-1}) 
		\psi_{\alpha}(g, h_e) \psi_{\alpha}(g_e, g) 
	\; \md g \md h_e \md g_e }
	{\mathcal{N}(\alpha)}
\\
&=  \sum_{\mathcal{C, C'}}\alpha^{A(\mathcal{C})} \bar{\alpha}^{A(\mathcal{C}')} 
 	\int_{SU(2)}\!
		\chi_{1/2}(g_eh_e^{-1}) \\
		&\times \prod_{\mathcal{L}\in\mathcal{C}}\chi_{1/2}\left( h_e, g \right) 
		\prod_{\mathcal{L}'\in\mathcal{C}'}\chi_{1/2}\left(g_e, g  \right) 
	\; \md g \md h_e \md g_e
\end{align*}
We integrate over $g_e$. If $g_e \notin \mathcal{C}'$ the integral gives zero. Otherwise we have simply 
$ \int_{SU(2)}\! \chi_{1/2}(g_eh_e^{-1}) \chi_{1/2}\left(g_eh\right)\; \md g_e  = \frac{1}{2}\chi_{1/2}(hh_e)$ ; 
substitute $h_e$ for $g_e$ with a factor one half. Finally, denoting by $\mathcal{C}'_e$ a configuration of loops 
containing the link $e$
\begin{align}
\langle \hat{j}_e \rangle  
&= \frac{1}{2 \mathcal{N}(\alpha)}
	\sum_{\mathcal{C}, \mathcal{ C}'_e}
		\alpha^{A(\mathcal{C})} \bar{\alpha}^{A(\mathcal{C}'_e)}  \\
		&\prod_{\mathcal{L} \in \mathcal{C}, \mathcal{L}'_e \in \mathcal{ C'}_e}
			\underbrace{
				 \int_{SU(2)}\!
					\chi_{1/2}\left(\prod_{e\in \mathcal{L}}^{\rightarrow} g_e \right) 
					\chi_{1/2}\left(\prod_{e\in \mathcal{L}'_e}^{\rightarrow} g_e \right) 
				\; \md g_e}
			_{=0  \text{ unless } \mathcal{L}=\mathcal{L}'_e} \nonumber \\
&= \frac{1}{2\mathcal{N}(\alpha)} 
	\sum_{\mathcal{ C}_e}
		 |\alpha|^{A(2\mathcal{C}_e)}  
 = \frac{\left|\alpha\right|^{2}}{\left( 1 + \left|\alpha\right|^{2} \right)^2}
\end{align}
The explicit evaluation of $\langle \hat{j}_e \hat{j}_{e'} \rangle$ follows the same 
steps. Distinguishing the two cases when the spins belong to the same loop or not, 
see FIG.\ref{correlations_example}, we have respectively 
$
\langle \hat{j}_e \hat{j}_{e'} \rangle
= \frac{1}{4} \frac{\left|\alpha\right|^{2}}{\left( 1 + \left|\alpha\right|^{2} \right)^2}  
$ 
and 
$
\langle \hat{j}_e \hat{j}_{e'} \rangle
=\frac{\left|\alpha\right|^{4}}{\left( 1 + \left|\alpha\right|^{2} \right)^4} 
$. 
In both cases, the correlation $\langle \hat{j}_e \hat{j}_{e'} \rangle = 
\langle \hat{j}_e \rangle  \langle \hat{j}_{e'} \rangle  $  is not in any way a function 
of the distance between the edges which is particularly clear when the edges don't 
belong to the same loop where the correlation is strictly zero. 

From the structure of state, we should have naively expected the 
correlations to scale in some way as the graph distance between the edges 
$e$ and $e'$.  This is in fact not the case since the averages counts every loops
meeting the edges in a democratic way (both configuration in FIG.\ref{correlations_example}
give the same correlations). 
Introducing a contribution to the amplitude proportional for instance to the number of loops 
can be a solution  to this issue. The limit $\beta= \gamma = 1$ has thus to be reconsidered 
to account for non trivial correlations.

\subsection{Example}

\begin{figure}[h]
  \centering

\begin{tikzpicture}[domain=-2.2:2.2]

\tikzstyle{dot}=[draw,circle,minimum size=2pt,inner sep=0pt,outer sep=0pt,fill=black]

	\def\a{1}
	\def\nbh{1}
	\def\nbw{2}
	
	\draw[blue] (\a,0) -- (0,0) -- (0,\a) -- (\a,\a) ;
	\node at (-0.3*\a,0.5*\a) {\textcolor{blue}{$h_S$}};

	\draw (\a,0) -- (2*\a,0) -- (2*\a,\a) -- (\a,\a) ;
	\node at (2.3*\a,0.5*\a) {$h_E$};
	
	\draw[red, thick] (\a,0) -- (\a,\a) ;
	\node at (1.2*\a,0.5*\a) {$h_b$};

	\foreach \i in {0,...,\nbw} {
		\foreach \j in {0,...,\nbh} {
					\coordinate[dot, black] () at (\i*\a,\j*\a);
		}
	
}

\end{tikzpicture}

  \caption{Illustrative example for the evaluation of the entanglement entropy for a two 
		loops state.}
  \label{example}
\end{figure}
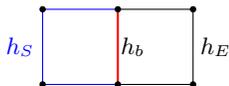

As an illustrative example, consider a two loops state whose wave function is 
$\psi(h_S,h_b,h_E) = 1 + \alpha\chi(h_S h_b) + \alpha \chi(h_E h_b^{-1}) + \alpha^2 \chi(h_S h_E)$. 
The reduced density matrix, obtained by taking two copies of the state and tracing out over
the environment has the form
\begin{align}
&\rho_{S}(h_b,h_S, h_b',h'_S) = \int \psi^*(h'_S,h'_b,h_E) \psi(h_S,b,h_E) \: \md h_E \nonumber \\
&= 1 + \alpha \chi(h_S h_b) + \overline{\alpha} \chi(h'_S h_b') + |\alpha|^2 \chi(h_S h_b)\chi(h'_S h_b') \nonumber \\
&+\frac{|\alpha|^2}{2} 
	\Big{[} \chi(h_bh_b'^{-1}) + \alpha \chi(h_bh'^{-1}_S) + \overline{\alpha} \chi(h_S h_b'^{-1}) \nonumber \\
&\qquad\qquad	+ |\alpha|^2 \chi(h_Sh'^{-1}_S)
	\Big{]}
\end{align}
The computation of the successive power of the reduced density matrix and 
the trace in then straightforward. We have $\trace{\rho_{S}^n (g,g')} = 
\frac{\left( 1 + |\alpha|^2 \right)^n}{\mathcal{N}(\alpha)} 
\left( 1 + \frac{|\alpha|^{2n}}{4^{n-1}} \right)$ and the entanglement
entropy follows formula \eqref{entropy}.

\section{Finding non trivial correlations}
\label{correlations}

\subsection{Distinguishing loops}

We saw in the last section why correlations were trivial for the restricted state
studied for entanglement entropy. This was coming from the fact that their 
was non distinction between loops passing through both links or not, see Figure 
\ref{correlations_example}. To understand how the solution comes about, 
let's look first at a simpler state constructed as the superposition of 
single loop holonomy
\begin{align}
\label{oneloopstate}
\psi(g_e) = \sum_{\mathcal{L}} \alpha ^{A(\mathcal{L})}
			\chi_{1/2}\left(\prod_{e\in \mathcal{L}}^{\rightarrow} g_e \right) 
\end{align}
This term is the first non trivial term of \ref{def_state} in an expansion of 
$\beta$ (for $\gamma =1$). The spin two points correlation function is 
straightforwardly evaluated as 
\begin{align}
\langle \hat{j}_e \hat{j}_{e'} \rangle &= 
	\int 
		\chi_{1/2}\left( g_e h_e^{-1} \right) \chi_{1/2}\left( g_{e'} h_{e'}^{-1} \right)	 
		\nonumber \\
		&\psi(h_e,h_{e'},g) \psi(g_e,g_{e'},g) \; \md g \md h_{e,e'} \md g_{e,e'} \nonumber \\
	&= \frac{1}{4} \frac{\mathcal{N}(e,e')}{\mathcal{N}}
\end{align}
with $\mathcal{N}(e,e') = \sum_{\mathcal{L}_{ee'}}|\alpha|^{A(\mathcal{L}_{ee'})}$ is a sum 
over all loops passing through both edge $e$ and $e'$. Now in this case, the correlations 
will scale non trivially on the minimal area between the edges since we must consider loops 
passing through both links at the same time. Here is the main difference between this state
and the previous one. 

We can go even further and analyze the entanglement entropy of a local region for this
state. In fact, the computation is completely similar to the one presented in  
\ref{entanglement}. However, the entanglement entropy doesn't follow an area law,
doesn't even scale as a function of the boundary degrees of freedom. Indeed, the 
area scaling came from the term 
$\ln\left( \frac{\mathcal{N}}{\mathcal{N}_S\mathcal{N}_E} \right)$ and 
the multiplicative nature of $\mathcal{N} = \mathcal{N}_S\mathcal{N}_E\mathcal{N}_{SE}$
whereas for \ref{oneloopstate}, $\mathcal{N}$ is additive. Thus the same 
contribution $\ln\left( \frac{\mathcal{N}}{\mathcal{N}_S\mathcal{N}_E} \right)$ is not 
only a function of the boundary degrees of freedom.

\subsection{The proposal}

The previous discussions show that two ingredients are necessary to 
obtain states with non trivial correlations and an entanglement entropy 
for a localized region to scale as the area of the boundary (at least to be 
a function of boundary degrees of freedom). Area law entanglement entropy
came from the loop structure of the toric code model, more precisely
all configurations of non intersecting and overlapping loops enter the superposition.
Non trivial correlations came on the other hand from the fact that a clear distinction 
between loops passing by both edges $e$ and $e'$ and those that do not was 
made. The solution presents itself when we come back to our original state 
\eqref{def_state} with amplitude scaling as the area and the number of 
loops (we omit the perimeter contribution since it can obstruct coarse-graining 
invariance),
 %
%
 \begin{align}
 \label{proposal}
\psi_{\alpha, \beta }(g_e) &=  
	\sum_{\mathcal{C}}\alpha^{A(\mathcal{C})}
		\prod_{\mathcal{L}\in\mathcal{C}} 
			\left( \beta \chi_{1/2}\left(\prod_{e\in \mathcal{L}}^{\rightarrow} g_e \right) \right)
\end{align}
The two requirements are here met.
The $\beta$ factor corresponds exactly to a number of loops contribution. It is
straightforward to generalize the following discussion to an arbitrary superposition
of holonomies $f(g) = \sum_{j=1/2}^{\infty} p_j \chi_j(g) $; the formal expressions
remains the same as before. 

Let's review its features. Concerning correlations, we can now distinguish 
a dominant term in the two points function. Indeed, what 
rendered the correlations topological initially was that their was no
distinction between the cases when the edges $e$ and $e'$ belong to the same loop
or different ones. With the additional $\beta$ contribution, we can now pinpoint a 
dominant term which is the one with minimal area and only one loop connecting the edges.
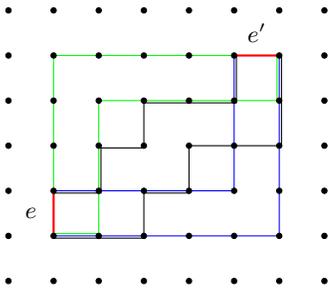
\begin{figure}[h]
  \centering

\begin{tikzpicture}[domain=-2.2:2.2]

\tikzstyle{dot}=[draw,circle,minimum size=2pt,inner sep=0pt,outer sep=0pt,fill=black]

	\def\a{.6}
	\def\nbh{6}
	\def\nbw{7}
	\pgfmathsetmacro{\l}{0.5*\a}
	
	\draw[red,thick] (5*\a,5*\a) -- ( 6*\a,5*\a) ;
	\node at (5.5*\a,5.5*\a) {$e'$};

	\draw[red, thick] ( \a,\a) -- (\a,2*\a) ;
	\node at (0.5*\a,1.5*\a) {$e$};
	
	\draw[blue] (\a,\a) -- (6*\a,\a) -- (6*\a,5*\a);
	\draw[blue] (\a,2*\a) -- (5*\a, 2*\a) -- (5*\a,5*\a);
	
	\draw[green] (\a,2*\a) -- (\a,5*\a) -- (5*\a,5*\a) ;
	\draw[green] (\a,1.05*\a) -- (2*\a,1.05*\a) -- (2*\a, 4*\a) -- (5.95*\a,4*\a) -- (5.95*\a, 5*\a) ;
	
	\draw (\a, 0.95*\a) -- (3*\a,0.95*\a) -- (3*\a,1.95*\a) -- (4*\a, 1.95*\a) -- (4*\a,3*\a) -- (6.05*\a,3*\a) -- (6.05*\a,5*\a);
	\draw (\a,1.95*\a) -- (2.05*\a, 1.95*\a) -- (2.05*\a, 2.95*\a) -- (3*\a,2.95*\a) -- (3*\a, 3.95*\a) -- 
			(5.05*\a, 3.95*\a) -- (5.05*\a, 5*\a);

	\foreach \i in {0,...,\nbw} {
		\foreach \j in {0,...,\nbh} {
					\coordinate[dot, black] () at (\i*\a,\j*\a);
		}
	
}

\end{tikzpicture}

  \caption{Set of possible minimal paths (not all drawn) joining the edges $e$ and $e'$ in red.}
  \label{distance}
\end{figure}
Denoting by $L_x$ and $L_y$ the horizontal and vertical graph distance respectively
connecting two given links $e$ and $e'$, the number of such minimal loops is ${A_{\text{min}} \choose L_y }$.
Thus, the dominant contribution to the spin correlations is
\begin{align}
\langle \hat{j}_e \hat{j}_{e'} \rangle 
	&= \frac{1}{4} |\beta| ^2 |\alpha|^{2A_{\text{min}}}
		{A_{\text{min}} \choose L_y } + o(|\beta| ^2 ,|\alpha|^{2A_{\text{min}}}) \\
	&\underset{N \rightarrow +\infty}{=} \frac{1}{4} |\beta| ^2  (2|\alpha|^2)^{A_{\text{min}}}
		\frac{\me^{-\frac{(L_x-L_y)^2}{2A_{\text{min}}}}}{\sqrt{A_{\text{min}}\pi/2}}
\end{align}
The correlations are now non topological. The correlations are maximum when the number 
of minimal paths joining the edges is maximal. This can be seen as entropic competition 
between the number of paths linking the edges $e$ and $e'$ and an energetic term 
$|\alpha|^{2A_{\text{min}}}$. The more connected the edges are the more correlated
they are. In the light of the distance from correlation point of view 
\cite{Livine_Terno_2005, Feller_2015}, the edges get closer when more 
different minimal paths of the graph exist.

The entropy can be obtained following the same steps as in Section \ref{entanglement}
by computing the successive power of the reduced density matrix and by employing 
the replica formula for the Von Neumann entropy. We have in the end an entanglement 
entropy function only of the boundary degrees of freedom, 
\begin{align}
S &= \ln(\mathcal{N}_{SE}(\alpha,\beta))  
- \frac{1}{\mathcal{N}(\alpha,\beta)}
	\sum_{\mathcal{C}_{SE}} 
		\left[ A(\mathcal{C}_{SE}) \ln\left(|\alpha^2|\right) \right. \nonumber \\
		&+ 
		\left. N(\mathcal{C}_{SE})\ln\left(\frac{|\beta|^2}{4}\right) \right]
		|\alpha|^{2A(\mathcal{C}_{SE})}|\beta|^{2N(\mathcal{C}_{SE})}
\end{align}
In the special case where $\alpha=1$ and $|\beta|=2$, we have a very simple 
expression for the entanglement entropy, 
\begin{align}
S = \ln(\mathcal{N}_{SE}(\alpha,\beta))
\end{align}
So in the end, we see that the state \eqref{proposal} is a good candidate 
to be a physical state, at least mirrors some features the true physical 
solution of the Hamiltonian constraint might be, since it as correlations 
that are function of some measure of distance in the graph through the 
minimal area and as an area law scaling entanglement entropy.

\section{Conclusion}
\label{conclusion}

In this paper, we introduced a class of states in loop quantum gravity whose
entanglement entropy for a bounded region scales as the area of the boundary 
(number of degrees of freedom)  and whose correlation functions between distant spins are
non trivial. Its structure is motivated by a condensed matter model, Kitaev's toric code 
model, where the ground state can be seen as a gas of loops on the lattice. Our 
ansatz mimics this structure, being defined as a superposition of non intersecting loops 
of arbitrary size. To each configuration, an amplitude function of the area, the perimeter 
or the number of loops is considered.

We showed that indeed the entanglement entropy of a region scales as the area of its boundary
using the replica trick method. The source of entanglement is seen to be exclusively due to  
loops crossing the boundary and the fact that the entanglement depends only on 
boundary degrees of freedom depends on the configuration structure. This analysis
serves also to illustrate extended Hilbert space ideas coming from research on local 
subsystems in gauge and gravity theories by seeing it as a clever way to purify a state.
On the side of the correlations, their non triviality come from 
the fact that some loops are distinct from the other . What's more, we showed that 
correlations grow as the number of minimal path joining the two spins is larger. 

The idea behind those kind of investigations is to have a clearer understanding
of the physical states of quantum gravity solving, ideally, all the constraints of 
the theory. From there we could infer the structure of the Hamiltonian constraints 
they are solution of pointing then toward the structure of the true quantum 
Hamiltonian constraint. Indeed, constructing a good Hamiltonian constraint 
is still under active research and we expect those retro-engineering studies
to open new perspectives.

At the end of the day, the goal would be to weave the standard loop quantum gravity techniques for designing quantum states of geometry by the action of holonomy operators and volume excitations with the MERA vision of local unitaries and (dis-)entangling operations, in order to understand the structure of (local) holographic states in (loop) quantum gravity.

\bibliographystyle{bib-style}
\bibliography{gravitation,mat_cond}


\appendix

\begin{widetext}	

\section{A first naive approach}
\label{naive}

We start our analysis of correlations on the simplest state on an oriented finite size regular
square lattice. This state $\ket{\psi}$ called in the following loop state is defined through
its wave-function in the following way:

\begin{align}
\psi(g_e) = \prod_p \chi_{1/2}\left(\prod_{e\in p}^{\rightarrow} g_e \right) 
			\chi_{1/2}\left(\prod_{e\in \partial}^{\rightarrow} g_e \right)
\end{align}
This state is constructed with spin $1/2$ holonomies of each loop which are $n_p$ in number 
in the bulk and a boundary term. Each spin of the bulk, living on a link of the graph, belongs
 to two holonomies. Only spin $0$ and $1$ are thus allowed in the bulk. On the contrary,
 a boundary spin belongs to only one holonomy and thus can only be a spin $1/2$.
 Up to now, this state is unnormalized. Its norm is  
$\langle \psi | \psi \rangle = 1 + \frac{1}{3^{n_p-1}}$. As the number
 of loops tends to infinity, the state becomes normalized.

Let us now study the correlations structure of this state. First of all, we study the
holonomy correlations. The average value of the holonomy $\chi_{j}(g_p)$ over the loop state is 

\begin{align}
\langle\chi_j(g_p)\rangle 
 &= \frac{\delta_{j2}\frac{1}{3^{n_p-1}} + \delta_{j1}\left( 1 + \frac{2}{3^{n_p-1}} \right) + 
\delta_{j0}\left( 1 + \frac{1}{3^{n_p-1}} \right)}{\langle \psi | \psi \rangle} \nonumber \\
&\underset{n_p \rightarrow +\infty}{\rightarrow}\delta_{j0} + \delta_{j1}
\end{align}
As the number of loops tends to infinity, the average value of the holonomy is one for spin
$1$ and $0$, zero for any other value which is quite pleasing following the structure of 
the state. For the correlations, we have to distinguish between adjacent loops and distant loops. 
The large $n_p$ limit reads

\begin{align}
\langle \chi_j(g_p) \chi_k(g_{p'}) \rangle  = \left\{
  \begin{array}{l l}
     \langle\chi_j(g_p)\rangle \langle\chi_k(g_{p'})\rangle  & \quad \text{if } \langle p, p' \rangle \\
    \left( 2\delta_{j0} + \delta_{j1}\right) \left( 2\delta_{k0} + \delta_{k1} \right)& \quad \text{if not}
  \end{array} \right.
\end{align}
Holonomies of distant loops are thus completely decorrelated. For adjacent loops, the two points
function is slightly modified. The loop state appears to be completely topological with respect to 
holonomies.

We proceed by the analysis of the spin correlation function. First of all, we precise the action of 
the spin operator of a given link $e$ on a given wave function $\psi(g)$. Thanks to the Peter-Weyl 
theorem, we have, denoting by $d_{j_E} = 2j_E + 1$,  
 
\begin{align}
\left( \hat{j}_e \psi\right) (g,g_e) 
= \sum_{j_e} d_{j_e}j_e \int_{SU(2)}\! \chi_{j_e}(g_e h^{-1})\psi(g,h) \md h
\end{align}
For the particular state we are considering, only the spin $0$ and $1$ contribute. For the average
value of the spin on a given link $e$, only the spin $1$ part gives a non zero contribution. 
In the large $n_p$ limit, we have 

\begin{align}
\langle \psi | \hat{j}_e | \psi \rangle  = \frac{3}{4}
\end{align}
This result can be understood quite easily as the probability to have a spin $1$ on the link $e$ , 
being the ratio between the dimension of the spin $1$ Hilbert space and the dimension of the 
total Hilbert space comprising spin $0$ and $1$. As for the two points function, we are exactly 
in the same situation as we were for the holonomies. We have

\begin{align}
\langle \psi | \hat{j}_e \hat{j}_{e'}| \psi \rangle = 
\langle \psi | \hat{j}_e | \psi \rangle\langle \psi | \hat{j}_{e'} | \psi \rangle
\end{align}
except for spins belonging to the same loop in which case the correlations are a simple constant 
for every situations. Once again the conclusion is that the loop state state is topological.

Thus, such a simple state is not a good candidate for our purpose. The problem seems to come 
from the fact that only spins belonging to the same loop see each other. This suggests the 
idea to construct a state with more extended loops and a more complex structure. Having 
also in mind that a state of the gravitational field should satisfy the holographic principle,
a particular model used in condensed matter and quantum information, the toric code model,
will allow us to shed some lights on a possible good loop structure.

\section{Proofs}
\label{proofs}

\subsection{Entanglement in toric code model}

\quote{
\textbf{Proposition}: \textit{The entanglement entropy associated to a 
given region $S$ whose (contractible) frontier possesses $n_{SE}$ degrees
of freedom in the ground state is $$S = n_{SE} - 1$$ }
}

\begin{proof}
The reduced density matrix of $S$ is obtained by tracing out 
the $E$ degrees of freedom. The most transparent way to 
do so is to really distinguish the loops belonging to $S$,
$E$ or both, as illustrated on fig.\ref{fig:SESE}. The respective sets are denoted 
$\mathcal{C}_S$, $ \mathcal{C}_E$ and $ \mathcal{C}_{SE}$
respectively. Then 

\begin{figure}[h]
  \centering

\begin{tikzpicture}[domain=-2.2:2.2]

\tikzstyle{dot}=[draw,circle,minimum size=2pt,inner sep=0pt,outer sep=0pt,fill=black]

	\def\a{.6}
	\def\nbh{5}
	\def\nbw{6}
	\pgfmathsetmacro{\l}{0.5*\a}
	
	\draw[blue, dashed] (1.5*\a,1.5*\a) -- (4.5*\a,1.5*\a) -- (4.5*\a,3.5*\a) -- (1.5*\a,3.5*\a) -- cycle ;
	\node at (4.5*\a,0.5*\a) {\textcolor{blue}{$\mathcal{SE}$}};

	\draw (0,0) -- (\a,0) -- (\a,\a) -- (0,\a) -- cycle;
	\node at (0.5*\a,4.5*\a) {$\mathcal{E}$};
	
	\draw[red, thick] (\a,4*\a) -- (\a,\a) -- (2*\a,\a) -- (2*\a,0) -- (3*\a,0) -- (3*\a,2*\a) -- (2*\a,2*\a) -- (2*\a,4*\a) -- cycle ;
	\draw[red, thick] (3*\a,4*\a) -- (5*\a,4*\a) -- (5*\a,2*\a) -- (4*\a,2*\a) -- (4*\a,3*\a) -- (3*\a,3*\a) -- cycle ;
	\node at (3.5*\a,2.5*\a) {$\mathcal{S}$};

	\foreach \i in {0,...,\nbw} {
		\foreach \j in {0,...,\nbh} {
					\coordinate[dot, black] () at (\i*\a,\j*\a);
		}
	
}

\end{tikzpicture}

  \caption{\label{fig:SESE}}
\end{figure}

\begin{align}
\rho_S
&= 	\sum_{\mathcal{C},\mathcal{C}'}
	\otimes_{\substack{
			\gamma\in\mathcal{C}\\
			\gamma'\in\mathcal{C}'
				}
		    }
			\ket{1_{e\in S}, 0_{e\not\in S}}
			\bra{1_{e'\in S}, 0_{e'\not\in S}} 
	\prod
		\langle1_{e\in E}, 0_{e\not\in E} | 
			1_{e'\in E}, 0_{e'\not\in E}\rangle
		\nonumber \\
&= \frac{4}{2^{n_p+1}}
		\left( \sum_{\mathcal{C}_E} 1 \right) 
		\sum_{\substack{\mathcal{C}_S, \mathcal{C}'_S \\ \mathcal{C}_{SE}}}
		\otimes_{\substack{
				\gamma\in\mathcal{C}_S \cup \mathcal{C}_{SE} \\
				\gamma'\in\mathcal{C}'_S\cup \mathcal{C}_{SE}
			    }
			    }
			\ket{1_{e\in\gamma}, 0_{e\not\in\gamma}}
			\bra{1_{e\in\gamma'}, 0_{e\not\in\gamma'}} 
		\nonumber \\
&= \frac{2^{n_E}}{2^{n_p-1}} 
		\sum_{\substack{
				\mathcal{C}_S \cup \mathcal{C}_{SE} \\
				\mathcal{C}'_S\cup \mathcal{C}_{SE}}}
		\otimes_{\substack{
				\gamma\in\mathcal{C}_S \cup \mathcal{C}_{SE} \\
				\gamma'\in\mathcal{C}'_S\cup \mathcal{C}_{SE}
			    }
			    }
			\ket{1_{e\in\gamma}, 0_{e\not\in\gamma}}
			\bra{1_{e\in\gamma'}, 0_{e\not\in\gamma'}}
\end{align}
The sets ($\mathcal{C}_S,\mathcal{C}_E,\mathcal{C}_{SE}$) are defined in the core of 
the paper. We first check that the trace of this density matrix is equal to one.

\begin{align}
\trace{\rho_S} = \frac{2^{n_E}}{2^{n_p-1}} 
			\left( 
				\sum_{\mathcal{C}_S, \mathcal{C}_{SE}} 1
			\right)
\end{align}
The counting is done by choosing $k$ plaquettes among $S$
or the boundary without choosing choosing the whole set 
of plaquette of the boundary (such a choice is equivalent to 
choosing loops that belong only to $S$ and/or $E$). Thus 
$\sum_{\mathcal{C}_S, \mathcal{C}_{SE}} 1 = 
\sum_k { n_S + n_{SE} -1 \choose k} = 2^{ n_S + n_{SE} -1}$. 
We this wa conclude that the reduced density matrix trace to 
unity. 

To determine the entropy, the method is to look at the 
squared density matrix. Explicitly we have 

\begin{align}
\rho_S^2
&= \left( \frac{2^{n_E}}{2^{n_p-1}} \right)^2
		\sum_{\substack{
				\mathcal{C}_S \cup \mathcal{C}_{SE} \\
				\mathcal{C}'_S\cup \mathcal{C}_{SE}
				      }
			}
		\sum_{\substack{
				\tilde{\gamma} \in \tilde{\mathcal{C}}_S 
					\cup \tilde{\mathcal{C}}_{SE} \\
				\tilde{\gamma} \in \tilde{\mathcal{C}}'_S 
					\cup \tilde{\mathcal{C}}_{SE}
				      }
			}
		\otimes_{\substack{ 
				\gamma\in\mathcal{C}_S \cup \mathcal{C}_{SE} \\
				\tilde{\gamma}' \in \tilde{\mathcal{C}}'_S 
					\cup \tilde{\mathcal{C}}_{SE}
					}
			}
			\ket{1_{e\in\gamma}, 0_{e\not\in\gamma}}
			\bra{1_{e\in\tilde{\gamma}'}, 0_{e\not\in\tilde{\gamma}'}}
		\underbrace
		{
		\prod_{\substack{ 
				\gamma'\in\mathcal{C}'_S\cup \mathcal{C}_{SE} \\
				\tilde{\gamma} \in \tilde{\mathcal{C}}_S 
					\cup \tilde{\mathcal{C}}_{SE}
				     } 
			}
			\langle \gamma'\in\mathcal{C}'_S \cup \mathcal{C}_{SE}  |
				 \gamma'\in\tilde{\mathcal{C}}_S \cup \tilde{\mathcal{C}}_{SE} 
			\rangle
		}_{\delta\left( \mathcal{C}'_S - \tilde{\mathcal{C}}_S \right)
		       \delta\left(\mathcal{C}_{SE} -\tilde{\mathcal{C}}_{SE}\right)}
		 \nonumber \\
&= \left( \frac{2^{n_E}}{2^{n_p-1}} \right)^2
		\left( \sum_{\mathcal{C}_S} 1 \right)
		\sum_{\substack{
				\mathcal{C}_S \cup \mathcal{C}_{SE} \\
				\mathcal{C}'_S\cup \mathcal{C}_{SE}
				      }
			}
		\otimes_{\substack{ 
				\gamma\in\mathcal{C}_S \cup \mathcal{C}_{SE} \\
				\gamma' \in \mathcal{C}'_S \cup \mathcal{C}_{SE}
					}
			}
			\ket{1_{e\in\gamma}, 0_{e\not\in\gamma}}
			\bra{1_{e\in\gamma'}, 0_{e\not\in\gamma'}}
= \frac{2^{n_E+n_S}}{2^{n_p-1}}\rho_S 
= \frac{1}{2^{n_{SE}-1}}\rho_S
\end{align}
\end{proof}

\subsection{Entanglement Kitaev state}
\label{Entanglement Kitaev state}

\begin{proof}
We consider a bipartite partition $S$, $E$ of the lattice and 
we want to evaluate the entanglement entropy between those
two regions. We'll make the assumption that the boundary of $S$
is given by contractible loop on the dual lattice for more simplicity.

The full density matrix reads 
\begin{align}
\rho_{SE}= \frac{1}{\mathcal{N}(\alpha)}
		\sum_{\mathcal{C}, \mathcal{C}'}
			\overline{\alpha}^{A(\mathcal{C}')} 
			 \alpha^{A(\mathcal{C})} 
		\prod_{\mathcal{L}\in\mathcal{C}, \mathcal{L}'\in\mathcal{C}'}
			\chi_{1/2}\left( \mathcal{L}\right)
			\chi_{1/2}\left( \mathcal{L}'\right)
\end{align}
with $\mathcal{N}(\alpha)$ the (squared) norm of the state
Once again, we distinguish the loops belonging to $S$, $E$ or 
both, writing respectively $\mathcal{C}_S$, $\mathcal{C}_E$
and $\mathcal{C}_{SE}$. The reduced density matrix for 
$S$ is obtained by tracing out the $E$ degrees of freedom.
So 
\begin{align}
\rho_{S} = \frac{1}{\mathcal{N}(\alpha)}
		\sum_{\mathcal{C},\mathcal{C}'}
			&\overline{\alpha}^{A(\mathcal{C}')} 
			 \alpha^{A(\mathcal{C})} 
		\prod_{\substack{
				\mathcal{L}_S, \mathcal{L}'_S
					}
			}
			\chi_{1/2}\left( \mathcal{L}_S\right)
			\chi_{1/2}\left( \mathcal{L}'_S\right) 
			\int
		\prod_{\substack{
				\mathcal{L}_{E}, \mathcal{L}'_{E}
					}
			}
			\chi_{1/2}\left( \mathcal{L}_{E} \right)
			\chi_{1/2}\left( \mathcal{L}'_{E} \right)
		\; \md g_{e\in  E}	\nonumber \\
			& \times  \int
		\prod_{\substack{
				\mathcal{L}_{SE}, \mathcal{L}'_{SE}
					}
			}
			\chi_{1/2}\left( \mathcal{L}_{SE} \right)
			\chi_{1/2}\left( \mathcal{L}'_{SE} \right)
		\; \md g_{e\in  E} \nonumber \\
\end{align}
This last integral is the one that creates entanglement between the 
bulk and the exterior region. The integration over $E$ gives rise to 
loops crossing the boundary of the region. The first integral can be done
straightforwardly and along with the area contribution gives an overall
factor from the environment $\mathcal{N}_E (\alpha)$.
We then have the final form of the reduced density matrix
\begin{align}
\rho_{S}(g,g') = \frac{\mathcal{N}_E (\alpha)}{\mathcal{N}(\alpha)} 
		\sum_{\substack{
				\mathcal{C}_S \cup \mathcal{C}_{SE} \\
				\mathcal{C}'_S \cup \mathcal{C}_{SE}
					}
			}
			\overline{\alpha}^{A(\mathcal{C}'_S \cup \mathcal{C}_{SE})} 
			 \alpha^{A(\mathcal{C}_S \cup \mathcal{C}_{SE})}
			 		\prod_{\substack{
				\mathcal{L}_S\in\mathcal{C}_S \\
				\mathcal{L}'_S\in \mathcal{C}'_S
					}
			}
			\chi_{1/2}\left( \mathcal{L}_S (g)\right)
			\chi_{1/2}\left( \mathcal{L}'_S (g')\right)
		\prod_{\mathcal{L}_{SE}\in\mathcal{C}_{SE}}
		\left(
		\frac{1}{2} 
			\chi_{1/2}\left( \mathcal{L}_{SE} (g,g') \right)
		\right)
\end{align}
To be more precise, $\mathcal{C}_{SE}$ is the set of loops in $S$
passing on (an even number of) punctures of the boundary. From the 
very construction of the state and the trace procedure the sets satisfy
$\mathcal{C}_S \cap \mathcal{C}_{SE} = \emptyset $ and 
$\mathcal{C}'_S \cap \mathcal{C}_{SE} = \emptyset $.
 
First of all, we can check that its trace is equal to one
\begin{align}
\trace{\rho_{S}} &= \frac{\mathcal{N}_E (\alpha)}{\mathcal{N}(\alpha)} 
		\sum_{\substack{
				\mathcal{C}_S \cup \mathcal{C}_{SE} \\
				\mathcal{C}'_S \cup \mathcal{C}_{SE}
					}
			}
			\overline{\alpha}^{A(\mathcal{C}'_S \cup \mathcal{C}_{SE})} 
			 \alpha^{A(\mathcal{C}_S \cup \mathcal{C}_{SE})}
			 		\prod_{\substack{
				\mathcal{L}_S\in\mathcal{C}_S \\
				\mathcal{L}'_S\in \mathcal{C}'_S
					}
			}
			\delta\left( \mathcal{L}_S - \mathcal{L}'_S \right)	
		\prod_{\mathcal{L}_{SE}\in\mathcal{C}_{SE}}
		\left(
		\frac{1}{2} 
			2
		\right) \nonumber \\
		&= \frac{\mathcal{N}_E (\alpha)}{\mathcal{N}(\alpha)}
 		\sum_{\mathcal{C}_S, \mathcal{C}_{SE}}
 			|\alpha|^{2A(\mathcal{C}_S)  + 2A(\mathcal{C}_{SE})} \nonumber \\
		&=\frac{\mathcal{N}_E (\alpha)}{\mathcal{N}(\alpha)}
			\sum_{k=0}^{n_S + n_{SE}} { n_S + n_{SE} \choose k}
		= 1
\end{align}

In order to calculate the entropy, we use the replica trick. We need then
to evaluate $\trace{\rho_S^n}$. Let's look at the square of the reduced density
matrix first, 
\begin{align}
\rho_{S}^2 (g,g') &= \left( \frac{\mathcal{N}_E (\alpha)}{\mathcal{N}(\alpha)}  \right)^2
		\sum_{\substack{
				\mathcal{C}_S \cup \mathcal{C}_{SE},
				\mathcal{C}'_S \cup \mathcal{C}_{SE} \\
				\tilde{\mathcal{C}}_S \cup \tilde{\mathcal{C}}_{SE},
				\tilde{\mathcal{C}}'_S \cup \tilde{\mathcal{C}}_{SE}
					}
			}
			\overline{\alpha}^{A(\mathcal{C}'_S \cup \mathcal{C}_{SE})} 
			 \alpha^{A(\mathcal{C}_S \cup \mathcal{C}_{SE})}
			 \overline{\alpha}^{A(\tilde{\mathcal{C}}'_S \cup \tilde{\mathcal{C}}_{SE})} 
			 \alpha^{A(\tilde{\mathcal{C}}_S \cup \tilde{\mathcal{C}}_{SE})} \nonumber \\
		&\times \prod
			\chi_{1/2}\left( \mathcal{L}_S (g)\right)
			\left[ \int
			\chi_{1/2}\left( \mathcal{L}'_S (h)\right)
			\chi_{1/2}\left( \tilde{\mathcal{L}}_S (h)\right)
			\; \md h \; \right]
			\chi_{1/2}\left( \tilde{\mathcal{L}}'_S (g')\right) \nonumber \\
		&\times
		\prod
		\left[
		\frac{1}{4} 
			\int \chi_{1/2}\left( \mathcal{L}_{SE} (g,h) \right)
			       \chi_{1/2}\left( \mathcal{L}_{SE} (h,g') \right)
			\; \md h
		\right] \nonumber \\
		&=\left( \frac{\mathcal{N}_E (\alpha)}{\mathcal{N}(\alpha)} \right)^2 
				\mathcal{N}_S (\alpha)
		\sum_{\substack{
				\mathcal{C}_S \cup \mathcal{C}_{SE} \\
				\mathcal{C}'_S \cup \mathcal{C}_{SE}
					}
			}
			\overline{\alpha}^{A(\mathcal{C}'_S \cup \mathcal{C}_{SE})} 
			 \alpha^{A(\mathcal{C}_S \cup \mathcal{C}_{SE})}
			 |\alpha|^{2A(\mathcal{C}_{SE})}
			 		\prod_{\substack{
				\mathcal{L}_S\in\mathcal{C}_S \\
				\mathcal{L}'_S\in \mathcal{C}'_S
					}
			}
			\chi_{1/2}\left( \mathcal{L}_S (g)\right)
			\chi_{1/2}\left( \mathcal{L}'_S (g')\right) \nonumber \\
		&\times \prod_{\mathcal{L}_{SE}\in\mathcal{C}_{SE}}
		\left(
		\frac{1}{2^3} 
			\chi_{1/2}\left( \mathcal{L}_{SE} (g,g') \right)
		\right)
\end{align}
By recursion we then obtain simply 
\begin{align}
\rho_{S}^n (g,g')&=\left( \frac{\mathcal{N}_E (\alpha)}{\mathcal{N}(\alpha)} \right)^{n}
				\mathcal{N}^{n-1}_S (\alpha)
		\sum_{\substack{
				\mathcal{C}_S \cup \mathcal{C}_{SE} \\
				\mathcal{C}'_S \cup \mathcal{C}_{SE}
					}
			}
			\overline{\alpha}^{A(\mathcal{C}'_S \cup \mathcal{C}_{SE})} 
			 \alpha^{A(\mathcal{C}_S \cup \mathcal{C}_{SE})}
			 |\alpha|^{2(n-1)A(\mathcal{C}_{SE})}
			 		\prod_{\substack{
				\mathcal{L}_S\in\mathcal{C}_S \\
				\mathcal{L}'_S\in \mathcal{C}'_S
					}
			}
			\chi_{1/2}\left( \mathcal{L}_S (g)\right)
			\chi_{1/2}\left( \mathcal{L}'_S (g')\right) \nonumber \\
		&\times \prod_{\mathcal{L}_{SE}\in\mathcal{C}_{SE}}
		\left(
		\frac{1}{2^{2n-1}} 
			\chi_{1/2}\left( \mathcal{L}_{SE} (g,g') \right)
		\right) \\
\trace{\rho_{S}^n (g,g')}&=\left( \frac{\mathcal{N}_E (\alpha)}{\mathcal{N}(\alpha)} \right)^{n}
				\mathcal{N}^{n-1}_S (\alpha)
		\sum_{\mathcal{C}_S \cup \mathcal{C}_{SE}} 
			\frac{|\alpha^{A(\mathcal{C}_S \cup \mathcal{C}_{SE})}|^{2n}}
				{\left(4^{n-1}\right)^{\#\mathcal{C}_{SE}}}
\end{align}
Now using the replica trick 
$S = - \left.\frac{\partial \trace{\rho_\cS^n}}{\partial n}\right|_{n=1}$
the entanglement entropy of the system $S$ is 

\begin{align}
S_{\alpha} =
	n_{SE} f(|\alpha|^2) +
	\frac{2\ln 2}{\left( 1 + |\alpha|^2 \right)^{n_{SE}}} 
		\sum_{\mathcal{C}_{SE}}\#\mathcal{L}_{SE} |\alpha|^{2A(\mathcal{L}_{SE})} 
\end{align}
with $f(|\alpha|^2) = \ln\left( 1 + |\alpha|^2 \right) - 
\frac{|\alpha|^2 }{ 1+ |\alpha|^2 } \ln\left( |\alpha|^2 \right) $.

\end{proof}
 
\end{widetext}
\end{document}